\documentclass[sigconf]{acmart}

\usepackage{booktabs} 

\usepackage{MnSymbol}
\usepackage{graphicx}
\usepackage{enumitem}
\usepackage{framed}
\usepackage{graphicx}
\usepackage{color}
\usepackage{array}  
\usepackage[]{hyperref} 
\usepackage{ragged2e}
\usepackage{enumitem}
\usepackage[colorinlistoftodos,prependcaption]{todonotes}
\usepackage[pagewise]{lineno} 
\hypersetup{
    colorlinks,
    citecolor=blue,
    filecolor=blue,
    linkcolor=blue,
    urlcolor=blue
}

\newcolumntype{L}[1]{>{\raggedright\let\newline\\\arraybackslash\hspace{0pt}}m{#1}}
\newcolumntype{C}[1]{>{\centering\let\newline\\\arraybackslash\hspace{0pt}}m{#1}}

\newcounter{EKXCommentsCounter}

\newcounter{TGOCommentsCounter}

\copyrightyear{2018}
\acmYear{2018}
\setcopyright{acmcopyright}
\acmConference[ICSE-SEIP '18]{40th International Conference on Software Engineering: Software Engineering in Practice Track}{May 27-June 3, 2018}{Gothenburg, Sweden}
\acmBooktitle{ICSE-SEIP '18: 40th International Conference on Software Engineering: Software Engineering in Practice Track, May 27-June 3, 2018, Gothenburg, Sweden}
\acmPrice{15.00}
\acmDOI{10.1145/3183519.3183539}
\acmISBN{978-1-4503-5659-6/18/05}

\begin{document}

\title{Exploration of Technical Debt in Start-ups}


\author{Eriks Klotins}
\authornote{Corresponding author}
\affiliation{%
  \institution{Blekinge Institute of Technology}
  \city{Karlskrona} 
  \state{Sweden} 
}
\email{eriks.klotins@bth.se}

\author{Michael Unterkalmsteiner}
\affiliation{%
  \institution{Blekinge Institute of Technology}
  \city{Karlskrona} 
  \state{Sweden} 
}

\author{Panagiota Chatzipetrou}
\affiliation{%
  \institution{Blekinge Institute of Technology}
  \city{Karlskrona} 
  \state{Sweden} 
}

\author{Tony Gorschek}
\affiliation{%
  \institution{Blekinge Institute of Technology}
  \city{Karlskrona} 
  \state{Sweden} 
}
                              
\author{Rafael Prikladnicki}
\affiliation{%
  \institution{Pontifical Catholic University of Rio Grande do Sul}
  \city{ Porto Alegre} 
  \state{Brazil} 
}

\author{Nirnaya Tripathi}
\affiliation{%
  \institution{University of Oulu}
  \city{Oulu} 
  \state{Finland} 
}

\author{Leandro Bento Pompermaier}
\affiliation{%
  \institution{Pontifical Catholic University of Rio Grande do Sul}
  \city{ Porto Alegre} 
  \state{Brazil} 
}

\renewcommand{\shortauthors}{E. Klotins et al.}

\begin{abstract}

\textit{Context:} Software start-ups are young companies aiming to build and 
market software-intensive products fast with little resources. Aiming to 
accelerate time-to-market, start-ups often opt for ad-hoc engineering 
practices, make shortcuts in product engineering, and accumulate technical debt.

\textit{Objective:} In this paper we explore to what extent precedents, 
dimensions and outcomes associated with technical debt are prevalent in 
start-ups.

\textit{Method:} We apply a case survey method to identify aspects of technical 
debt and contextual information characterizing the engineering context in start-ups.

\textit{Results:} By analyzing responses from 86 start-up cases we found that 
start-ups accumulate most technical debt in the testing dimension, despite 
attempts to automate testing. Furthermore, we found that start-up team size and 
experience is a leading precedent for accumulating technical debt: larger teams 
face more challenges in keeping the debt under control.

\textit{Conclusions:}
This study highlights the necessity to monitor levels of technical debt and to 
preemptively introduce practices to keep the debt under control. Adding more 
people to an already difficult to maintain product could amplify other 
precedents, such as resource shortages, communication issues and negatively 
affect decisions pertaining to the use of good engineering practices.


\end{abstract}

%
%



\maketitle

\section{Introduction}

Start-ups are important suppliers of innovation, new software products and 
services. However, engineering the software in start-ups is a complex endeavor 
as the start-up context poses unique challenges to software 
engineers~\cite{Giardino2015}. 
As a result of these challenges, most start-ups 
do not survive the first few years of operation and cease to exist before 
delivering any value~\cite{Blank2013,Giardino2014}.

Uncertainty, changing goals, limited human resources, extreme time and resource 
constraints are reported as characteristic to 
start-ups~\cite{Paternoster2014a, Giardino2015}. To cope with 
such forces, start-ups make a trade-off between internal product quality and faster time-to-market, favoring the latter. As a consequence, start-ups accumulate technical debt~\cite{Unterkalmsteiner}.

Technical debt is a metaphor to describe not quite right engineering 
solutions in a product that adds friction to its further development and 
maintenance.
The extra effort associated with this friction, i.e. the ``interest'', needs 
to be paid every time a sub-optimal solution is touched~\cite{Kruchten2012}.
Over time, the cumulative interest may exceed the effort needed to remove the 
debt, i.e. the ``principal''.
The compound effects of sub-optimal solutions can reduce
development team efficiency and overall quality. However, there is the 
belief 
among start-ups that any amount of 
technical debt can be written off if a feature or the whole product is not 
successful in the market~\cite{Unterkalmsteiner}.

The strategy to accumulate technical debt can backfire if a start-up survives 
long enough and fails to put its technical debt under control. An unstable and 
difficult to
maintain product adds risk to the company, for example, by limiting the ability 
to quickly enter into new markets (i.e. to pivot~\cite{Terho2015}) or to launch 
new innovative features~\cite{Tom2013,Klotinsc}. That said, we are not 
advocating for the removal of all technical debt. Rather, we are interested to 
see an overview of how technical debt influences start-ups and to enable 
start-up teams to make better decisions in regards to the trade-off between 
quality and time-to-market.

Technical debt has been extensively studied in the context of established 
companies and in relation to software maintenance~\cite{Li2015, Sjoberg2013}. For example, 
Tom et al.~\cite{Tom2013} present a taxonomy comprising precedents, 
dimensions, and outcomes of technical debt. We adopt the terminology from this taxonomy to enable traceability.

Precedents are contextual factors in the development organization that contribute to 
the accumulation of technical debt, e.g. a lack of resources. Dimensions 
describe different types of technical debt, e.g. documentation, architecture, 
or testing debt. Outcomes refer to consequences of having 
excess technical debt, such as impaired productivity or quality~\cite{Tom2013}.

While technical debt is a liability, development teams should manage it and use it as a leverage to attain otherwise unattainable goals~\cite{Tom2013}. In the start-up context, the concept of 
technical debt is explored only superficially. Giardino et 
al.~\cite{Unterkalmsteiner} 
argue that the need for speed, cutting 
edge technologies and uncertainty about a product's market potential are the 
main 
precedents for cutting corners in product engineering. However, if a start-up 
survives past its initial phases, management of technical debt becomes more and 
more important~\cite{Crowne2002,Unterkalmsteiner}.



Our earlier study on software engineering anti-patterns in 
start-ups~\cite{Klotinsc} indicates that poorly managed technical debt could be one 
contributing factor to high start-up failure rates, driven by poor product 
quality and difficult maintenance. Negative effects of technical debt on 
team productivity has also been observed~\cite{Unterkalmsteiner}.




In this study, we explore how start-ups estimate technical debt, what are 
precedents for accumulating technical debt, and to what extent start-ups 
experience outcomes associated with technical debt. We use a case survey as 
data source and apply a combination of quantitative and qualitative methods to 
explore technical debt in the surveyed companies. Our objective is to provide a 
fine-grained understanding of technical debt and its components that could 
provide a basis for defining start-up context-specific practices for 
technical debt management.

The main contribution of this paper is an empirical investigation that 
identifies the key precedents for the accumulation of technical debt in 
software start-ups, and the primary dimensions where the accumulation of debt 
has been observed by practitioners. 

The rest of the paper is structured as follows. In Section~\ref{sec_bgrw} we 
introduce relevant concepts to understand our study. Section~\ref{sec_rm} 
presents the study design while results are presented in 
Section~\ref{sec_results}. The results are discussed and interpreted in 
Section~\ref{sec_dis}. Section~\ref{sec_conclusions} concludes the paper.

\section{Background and related work}
\label{sec_bgrw}

\subsection{Software start-ups}
Software start-ups are small companies created for the purpose of developing 
and bringing an innovative product or service to market, and to benefit
from economy of scale. Even though start-ups 
share many characteristics with small and medium enterprises, start-ups are 
different due to the combination of challenges they face~\cite{Sutton2000,metzger2014software}. 

Start-ups are characterized by high risk, uncertainty, lack of resources, rapid 
evolution, immature teams, and time pressure among other factors. However, 
start-ups are flexible to adopt new engineering practices, and reactive to keep 
up with emerging technologies and markets~\cite{Giardino2014, Sutton2000}. 

Start-up companies rely on external funding to support their endeavors. In 2015 alone, start-up companies have received investments of 429 billion USD in the US and Europe alone~\cite{PitchBookData2015a,PitchBookData2015}. With an optimistic start-up failure rate of 75\% that constitutes of 322 billion USD of capital potentially wasted on building unsuccessful products. 

Earlier studies show that product engineering challenges and inadequacies in 
applied engineering practices could be linked to start-up 
failures~\cite{Giardino2015,Klotinsc}. To what extent software engineering 
practices are responsible or linked to success rate is very hard to judge. 
However, if improved software engineering practices could increase the 
likelihood of success by only a few percent, it would yield a significant impact 
on capital return.

\subsection{Technical debt}

Technical debt is a metaphor to describe the extra effort arising from 
maintaining and removing suboptimal or flawed solutions from a software product. 
Technical debt can be attributed to the software itself (e.g. source code), and also other artifacts and processes that comprise the product, and are relevant for maintenance and evolution of the product. For example, user manuals, knowledge distribution, operational processes, and infrastructure~\cite{Tom2013}.

Suboptimal 
solutions find their way into software products due to a variety of reasons, such as 
ignorance of good engineering practices, oversight, lack of skills, or 
pragmatism~\cite{alves2016identification}. 
Taking engineering shortcuts and delivering flawed solutions is 
often used as leverage to achieve faster time-to-market. However, the debt 
should be re-payed by removing flawed solutions from the 
product~\cite{Tom2013, Li2015}. 

When not addressed, suboptimal solutions make maintenance and evolution of software products difficult, any changes in the product require more effort than without the
debt. This extra effort takes time away from developing new features 
and may overwhelm a team with firefighting tasks just to keep the product 
running, and decreases product quality altogether~\cite{Kruchten2012}.

\begin{figure}[!t]
    \centering
    \includegraphics[width=\textwidth*4/9]{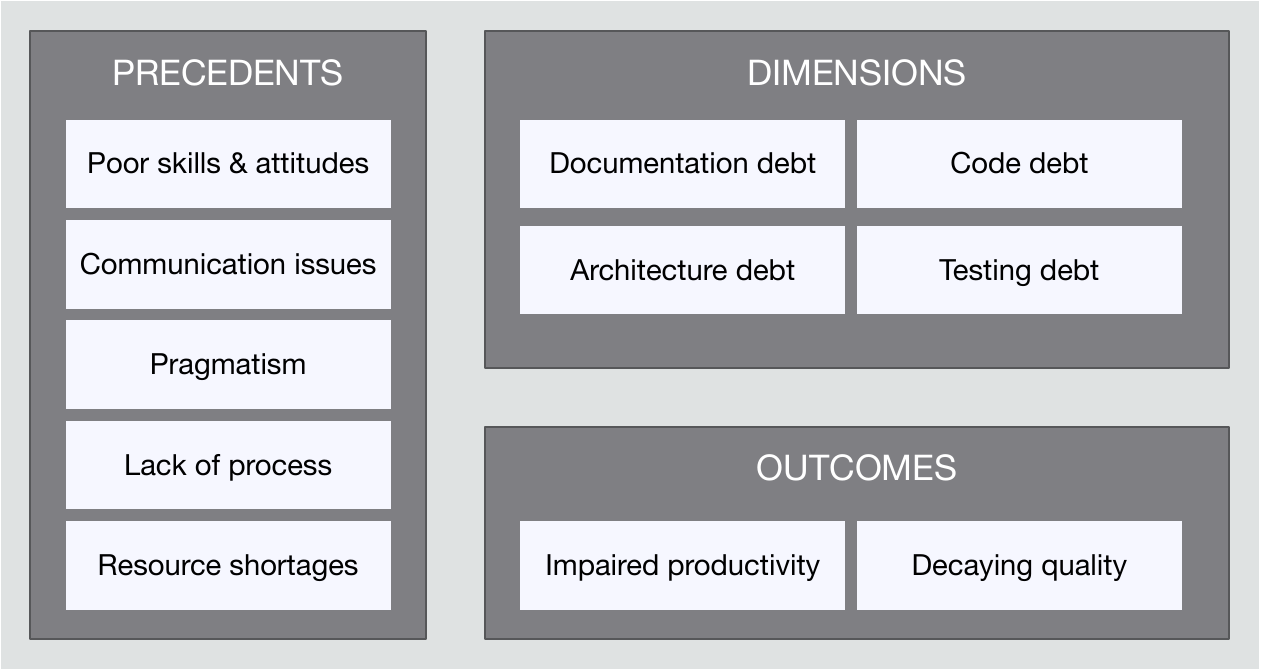}
    \caption{Aspects of technical debt}
    \label{fig_construct}
\end{figure}

Giardino et al.~\cite{Unterkalmsteiner} argue that technical debt in start-ups 
accumulates from prioritizing development speed over quality, team aspects, and 
lack of resources. We combine results from their work, which is specific to 
start-ups, with a general taxonomy of technical debt by Tom et 
al.~\cite{Tom2013}. We adopt the model of precedents, dimensions, and outcomes 
as proposed by Tom et al.~\cite{Tom2013} and map it with the categories of 
the Greenfield start-up model~\cite{Unterkalmsteiner} to identify and to focus 
on relevant aspects of technical debt for start-ups, see Fig.~\ref{fig_construct}.

As precedents, we study engineering skills and attitudes, communication 
issues, pragmatism, process, and resources. We explore technical debt in forms 
of code smells, software architecture, documentation, and testing. Furthermore, 
we attempt to understand to what extent team productivity and product quality 
is a challenge in start-ups. We use this conceptual model of technical debt as 
a basis to scope and define the research methodology, discussed next.

\section{Research methodology}
\label{sec_rm}


\subsection{Research questions}

To achieve our goal and to drive the study we formulate the following research questions:
\begin{itemize}[label={},leftmargin=0in]

\item \textbf{RQ1:} How do start-ups estimate technical debt?

\textbf{Rationale:} Technical debt can incur in different forms, for example, 
as code smells, incomplete or outdated documentation, suboptimal software 
architecture, or shortcuts taken in testing~\cite{Li2015}. We aim to understand 
how start-ups estimate different types of technical debt, what types of 
technical debt are prevalent in start-ups and primary candidates for further 
investigation. In addition, what types of technical debt are least accumulated, 
i.e. are irrelevant or already well managed in the start-up context.
\\
\item \textbf{RQ2:} What are precedents of technical debt in start-ups?

\textbf{Rationale:} Earlier studies report a number of precedents contributing 
to the accumulation of technical debt, such as prioritizing time-to-market over 
product quality and severe lack of resources~\cite{Unterkalmsteiner}, developer 
skills and attitude, lack of process, oversight, and ignorance~\cite{Tom2013}. 
We aim to corroborate what precedents, identified by earlier studies in other 
contexts, are also present in start-ups. 
\\
\item \textbf{RQ3:} What outcomes linked to technical debt do start-ups report?

\textbf{Rationale:} Decreasing productivity, decaying morale, product quality 
issues, and increasing risks are reported as outcomes of technical 
debt~\cite{Tom2013, Unterkalmsteiner}. Yet, there is a belief 
that any amount of technical debt can be written off if a product or a specific 
feature does not succeed in market~\cite{Unterkalmsteiner}. We aim to 
corroborate what outcomes, identified by earlier studies and linked to 
increased amounts of technical debt, do start-ups report.

\end{itemize}

\subsection{Data collection}

We used a case survey method to collect primary data from start-up 
companies~\cite{petersen2017choosing,Larsson1993}. 

The case survey method is based on a questionnaire and is a compromise between a traditional case study and 
a regular survey~\cite{Klotins2017}. We have designed the questionnaire to collect 
practitioners experiences about specific start-up cases.

During the questionnaire design phase, we conducted multiple internal and 
external reviews to ensure that all questions are relevant, clear and that we 
receive meaningful answers. First, the questions were reviewed in multiple 
rounds by the first three authors of this paper to refine scope of the survey 
and question formulations. Then, with help of other researchers from the 
Software Start-up Research Network\footnote{The Software Start-up Research 
Network, https://softwarestartups.org/}, we conducted a workshop to gain 
external input on the questionnaire. A total of 10 researchers participated and 
provided their input. 

Finally, the questionnaire was piloted with four practitioners from different 
start-ups. During the pilots, respondents filled in the questionnaire while 
discussing questions, their answers and any issues with the first author of 
this paper. 

As a result of these reviews, we improved the question formulations and removed 
some irrelevant questions. The finalized questionnaire\footnote{
\url{http://startupcontextmap.org/exp-survey/woifenw2}} contains 85 
questions in 10 sections. The questionnaire captures 285 variables from 
each start-up case.

From all the variables, 45 variables focus on capturing the magnitude of 
dimensions, precedents, and outcomes linked to technical debt\footnote{The subset of questions used in this study is available here: \url{http://eriksklotins.lv/uploads/TD-in-start-ups-questions.pdf} }. The questions
capture the respondents' agreement with a statement on a Likert scale: not at 
all (1), a little (2), somewhat (3), very much (4). 
The values indicate the degree of agreement with a statement. Statements are formulated consistently in a way that lower values indicate less precedents, less outcomes, and less technical debt. 

In addition to questions pertaining technical debt, the questionnaire contains questions 
inquiring the engineering context in the start-up and applied software 
engineering practices. 

The data collection took place between December 1, 2016, and June 15, 2017. The 
survey was promoted through personal contacts, by attending industry events, 
and by posts on social media websites. Moreover, we invited other researchers 
from the Software Start-up Research Network to collaborate on the data 
collection. This collaboration helped to spread the survey across many 
geographical locations in Europe, North and South America, and Asia.

\subsection{Data analysis}



To analyze the survey responses we used a number of techniques. We started by
screening the data and filtering out duplicate cases, responses with few questions 
answered, or otherwise unusable responses. In the screening we attempt to be as
inclusive as possible and do not remove any cases based on the provided responses.

The respondent estimates on technical debt aspects are measured on an ordinal scale, measured from 1 (not at all) to 4 (very much). Respondent and start-up demographics such as age and years of operation are measured with categorical variables on a nominal scale.

Overall, we analyze responses from 86 start-up cases, 75 data-points per each case, and 6450 data-points overall. To gain an overview of the data, the results were visualized by histograms, box-plots and contingency tables~\cite{haberman1973analysis}.

We use the Chi-Squared test of association to test if the associations between the examined variables are not due to chance. To prevent Type I errors, we used exact tests, specifically, the Monte-Carlo test of statistical significance based on 10~000 sampled tables and assuming $(p = 0.05)$~\cite{hope1968simplified}. 


To examine the strength of associations we use Cramer's V test. We interpret the test results as suggested by Cohen~\cite{cohan1988statistical}, see Table~\ref{table_cramersv}.
To explore specifics of the association, such as which cases are responsible 
for this association, we perform post-hoc testing using adjusted residuals. We 
consider an adjusted residual significant if the absolute value is above 1.96 
$(Adj.residual > 1.96)$, as suggested by 
Agresti~\cite{agresti1996introduction}. 
The adjusted residuals drive our analysis on how different groups of start-ups estimate aspects of technical debt. However, due to the exploratory nature of our study, we do not state any hypotheses upfront and drive our analysis with the research questions.

\begin{table}[!t]
\renewcommand{\arraystretch}{1.1}
\caption{Interpretation of Cramer's V test  }
\label{table_cramersv}
\centering

\begin{tabular}{C{1in}L{1.5in}}

Cramer's V value & Interpretation   \\
\hline
$\ge 0.1 $ & Weak association       \\
$\ge 0.3 $ & Moderate association   \\
$\ge 0.5 $ & Strong association     \\
\hline
\end{tabular}
\end{table}

Full results, contingency tables, histograms and calculation details are 
accessible 
on-line\footnote{\url{http://eriksklotins.lv/uploads/TD-in-start-ups-sm.pdf}} 
for a full disclosure.

\subsection{Validity threats}
In this section we follow guidelines by Runeson et al.~\cite{Runeson2012} and discuss four types of validity threats and applied countermeasures in the context of our study.

\subsubsection{Construct validity}
Construct validity concerns whether operational measures really represent the 
studied subject~\cite{Runeson2012}. A potential threat is that the statements 
we use to capture respondent estimates are not actually capturing the studied 
aspects of technical debt. 

To address this threat we organized a series of workshops with other 
researchers and potential respondents to ensure that questions are clear, to 
the point, and capture the studied phenomenon. 

Each aspect, i.e. type of precedent, is triangulated by capturing it by at 
least three different questions in the questionnaire. To avoid biases stemming 
from respondents opinions about technical debt and to capture the actual 
situation we avoid mentioning technical debt in the questions. Instead, we 
formulate the questions indirectly to capture respondent estimates on different 
aspects associated with technical debt. For example, we ask whether they find 
it difficult to understand requirements documentation. 

To accommodate for the fact that a respondent may not know answers to some of 
the questions, we provide an explicit ''I do not know'' answer option to all 
Likert scale questions. 

\subsubsection{Internal validity}
This type of validity threat addresses causal relationships in the study 
design~\cite{Runeson2012}. In our study we use a model of precedents, 
dimensions and outcomes of technical debt. The literature, for example, Tom et 
al.~\cite{Tom2013} and Li et al.~\cite{Li2015}, suggest that there is a 
causality between the three. We, however, present respondent estimates on 
precedents, dimensions and the outcomes separately without considering or 
implying any causality.

\subsubsection{External validity}
This type of validity threat concerns to what extent the results could be valid 
to start-ups outside the study~\cite{Runeson2012}. The study setting for 
participants was close to real life as possible, that is, the questionnaire was 
filled in without researcher intervention and in the participants own 
environment.

A sampling of participants is a concern to external validity. We use convenience 
sampling to recruit respondents and with help of other researchers, distributed 
the survey across a number of different start-up communities. Demographic 
information from respondent answers shows that our sample is skewed towards 
active companies, respondents with little experience in start-ups, young 
companies and small development teams of 1-8 engineers. In these aspects our sample fits the general characteristics of start-ups, see for 
example, Giardino et al.~\cite{Giardino2014, Giardino2015} and Klotins et 
al.~\cite{Klotins2015b}. However, there clearly is a 
survivorship bias, that is, failed start-ups are underrepresented, thus our 
results reflect state-of-practice in active start-ups. 

Another threat to external validity stems from case selection. The 
questionnaire was marketed to start-ups building software-intensive products, 
however due to the broad definition of software start-ups (see Giardino et 
al.~\cite{Giardino2014}), it is difficult to differentiate between start-ups 
and small medium enterprises. We opted to be as inclusive as possible and to 
discuss relevant demographic information along with our findings.

\subsubsection{Conclusion validity}

This type of validity threat concerns the possibility of incorrect 
interpretations arising from flaws in, for example, instrumentation, respondent 
and researcher personal biases, and external influences~\cite{Runeson2012}. 

To make sure that respondents interpret the questions in the intended way we 
conducted a number of pilots, workshops and improved the questionnaire 
afterwards. To minimize the risk of systematic errors, the calculations and 
statistical analysis was performed by the first and the third author 
independently, and findings were discussed among the authors.

To strengthen reliability and repeatability of our study, all survey materials and calculations with immediate results are published online.

\section{Results}
\label{sec_results}

\begin{figure}[ht]
    \centering
    \includegraphics[width=\textwidth * 4/9]{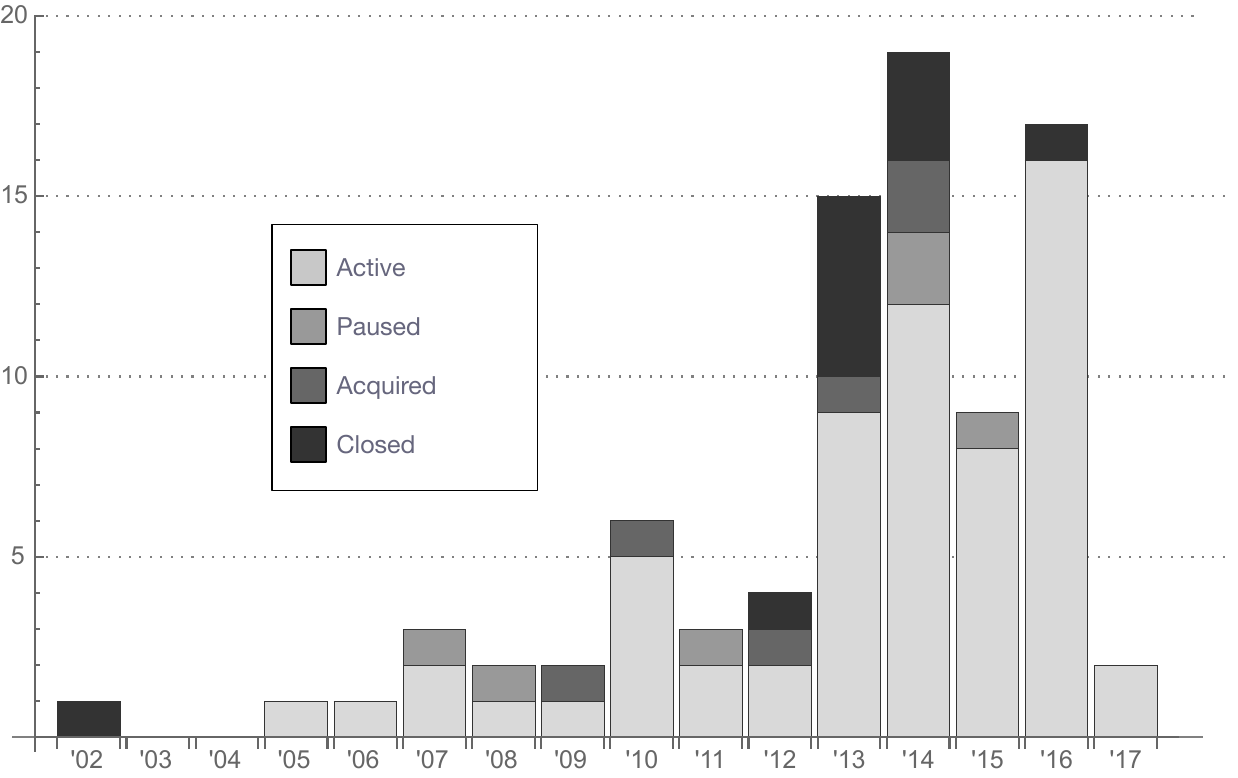}
    \caption{Distribution of start-ups by the founding year and their current state}
    \label{fig_FoundedxState}
\end{figure}

To answer our research questions we analyze 6450 data-points from 86 start-up cases. The majority of these start-ups (63 out of 86, 73\%) are active and had been operating for 1\,-\,5 years (58 out of 
86, 67\%), see Fig.~\ref{fig_FoundedxState}. 
Start-ups are geographically distributed among Europe (34 out of 86, 40\%), South America (41 
out of 86, 47\%), Asia (7 out of 86) and North America (2 out of 86).

\begin{figure}[ht]
    \centering
    \includegraphics[width=\textwidth * 4/9]{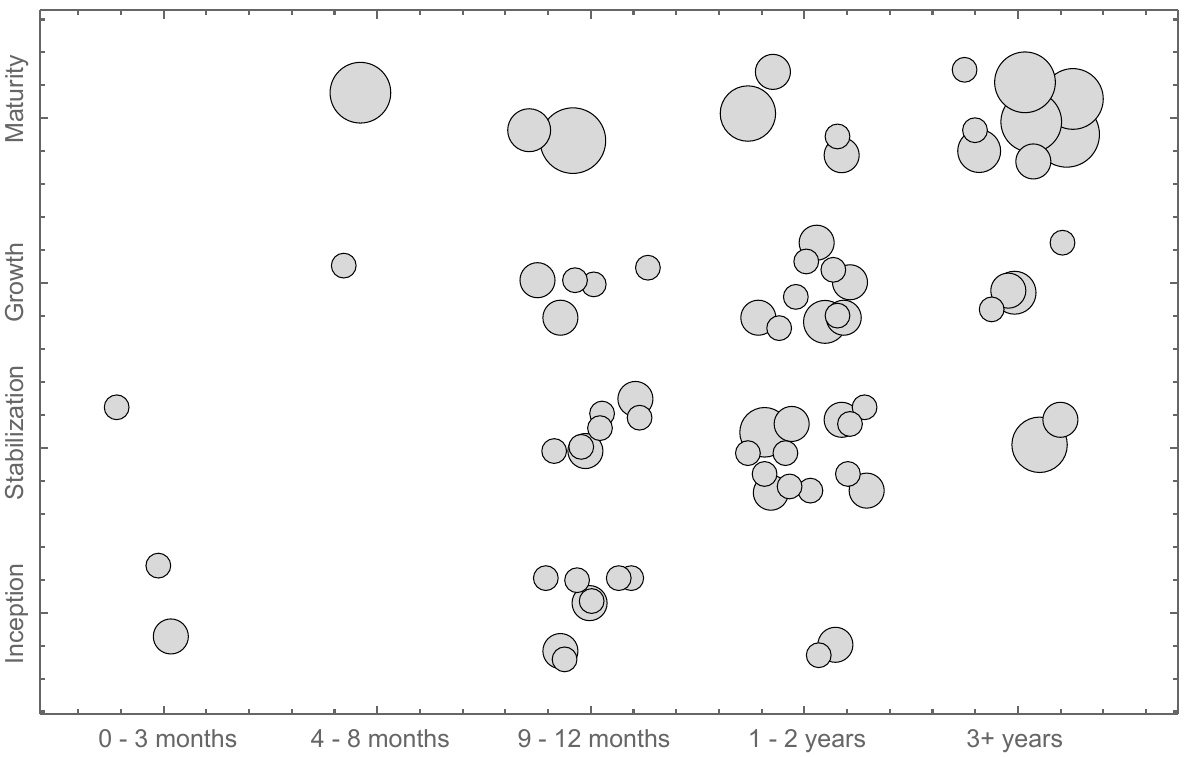}
    \caption{Distribution of start-ups by product phase and length of operation}
    \label{fig_StatexLong}
\end{figure}

Our sample is about equally distributed in terms of the product development 
phase. We follow a start-up life-cycle model proposed by 
Crowne~\cite{Crowne2002} and distinguish between inception, stabilization, 
growth and maturity phases. In our sample, 16 start-ups have been working on a 
product but haven't yet released it to market, 24 teams had released the first 
version and actively develop it further with customer input, 26 start-ups have 
a stable product and they focus on gaining customer base, and another 16 
start-ups have mature products and they focus on developing variations of their 
products. The distribution of start-ups by their life-cycle phase and length of 
operation is shown in Fig.~\ref{fig_StatexLong}. In the figure, bubble size 
denotes the
a number of people in the team. Most start-ups in the sample (75 out of 86, 87\%) have small teams 
of 1\,-\,8 engineers actively working on the product.

About an equal number of start-ups had indicated that they work on more than 
one product at the time. Start-ups in our sample do per-customer customization 
to some extent: 10 companies (11\%) had specified that they tailor each product 
instance to a specific customer, 30 companies (35\%) do not do per-customer 
customization at all, while 43 start-ups (49\%) occasionally perform product 
customization for an individual customer.

The questionnaire was filled in mostly by start-up founders (64 out of 86, 
74\%) and engineers employed by start-ups (15 out of 86, 17\%). About a half of 
respondents have specified that their area of expertise is software engineering (49 out of 86, 56\%). Others have specified marketing, their respective domain, and business development as their areas of expertise.

Respondents length of software engineering experience ranges from 6 months to 
more than 10 years. A large portion of respondents (44 out of 86, 51\%) had 
less than 6 months of experience in working with start-ups at the time when 
they joined their current start-up.


\subsection{Dimensions of technical debt}\label{sec:dimtd}

We start our exploration by looking at the extent to which the dimensions of 
technical debt (documentation, architecture, code, and testing) are present in 
the surveyed start-ups. 
We quantify the degree of technical debt by aggregating respondent answers on questions pertaining to each dimension,
 Answers were given on a Likert scale where higher values indicate more estimated technical debt in a given dimension. 

\begin{figure}[ht]
    \centering
    \includegraphics[width=\textwidth * 4/9]{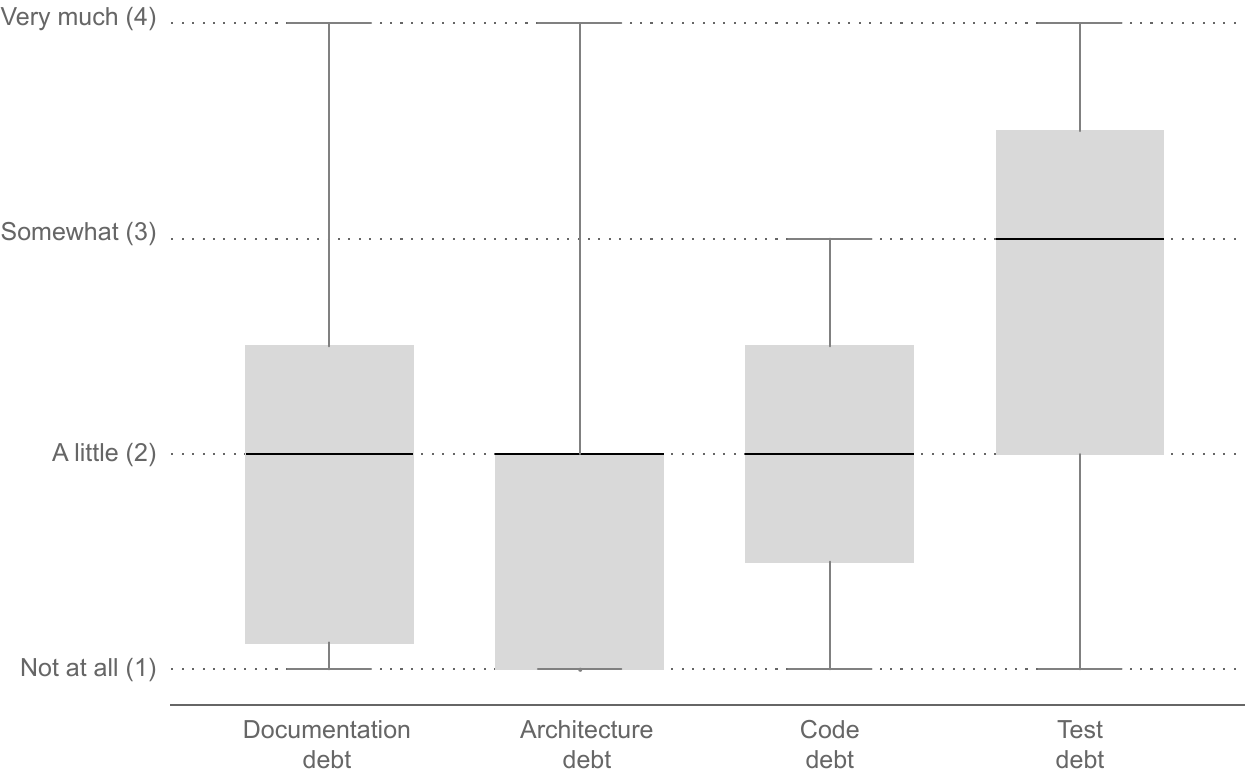}
    \caption{Estimates for the prevalence of different dimensions of technical 
    debt from the case survey}
    \label{fig_dimensions}
\end{figure}

Responses from the whole sample indicate that start-ups estimate some technical 
debt (2 on a scale from 1 to 4) in documentation, architecture, and code 
dimensions, while testing debt is estimated as the most prevalent (3 in a scale 
from 1 to 4). Fig.~\ref{fig_dimensions} shows the median 
(dark horizontal line), first and third quartile, and minimum and maximum 
estimates on all statements pertaining to a specific debt type.

To explore the estimated degree of technical debt further, we analyze the 
influence of respondent demographics, such as relationship with the start-up 
and background, and start-up demographics, such as product life-cycle phase, 
team skill level and longevity of the start-up, on the responses. 

\begin{table*}[!t]
\renewcommand{\arraystretch}{1.1}
\caption{Results of Cramer's V test on associations between dimensions and start-up characteristics with $p \less 0.05$ }
\label{table_characteristics_dimensions}
\centering

\begin{tabular}{C{0.2in}L{1.5in}L{0.8in}L{0.8in}L{0.8in}L{0.8in}L{1in}}

\# & Characteristic           & Documentation 
                                      & Architecture  
                                              & Code 
                                                      & Testing 
                                                              & All \\
\hline

1 & State of the start-up     & 0.346  & 0.326  & 0.414 & -      & 0.318  \\
2 & Product phase             & -      & 0.329  & -     & -      & -      \\
3 & Overall team size         & -      & -      & 0.427 & -      &-       \\
4 & Level of domain knowledge & 0.334  & -      & -     & -      &-       \\
5 & Per-customer tailoring    & -      & -      & 0.423 & -      & -      \\
6 & Practical experience      & 0.337  & -      & -     & -      & -      \\

\hline
\end{tabular}
\end{table*}

The analysis shows that only start-up state, that is if the start-up is active, 
paused, 
acquired, or closed, has an effect on the overall estimates of technical debt, 
see Table~\ref{table_characteristics_dimensions}. In the table we show strength 
(measured by Cramer's V test) of statistically significant associations ($p 
\less 0.05$, measured by Chi-Square test) between relevant characteristics of 
start-ups and technical debt dimensions. In the last column we show if the 
characteristic has an effect on all dimensions together. 

Observe that the level of engineering skills and domain knowledge pertains to the whole team. However, practical experience pertains only to the respondent. We show respondent characteristics as well to illustrate to what extent respondents background influences their responses. For instance, respondents with more practical experience estimate documentation debt more critically, see Table~\ref{table_characteristics_dimensions}, and are more critical about skills shortages, see Table~\ref{table_characteristics_precedents}.

We highlight important findings in framed boxes and discuss them in 
Section~\ref{sec_dis}.
\begin{framed}
\textbf{Finding 1:} Start-ups that are in the active category estimate technical debt, overall in all dimensions, lower than closed or acquired start-ups.
\end{framed}

Product phase, team size and level of domain knowledge have
effects on individual technical debt dimensions. We present these results next.

\subsubsection{Documentation debt}

Documentation debt refers to any shortcoming in documenting aspects of  
software development, such as architecture, requirements, and test 
cases~\cite{Li2015}.


We look into requirements, architecture and test documentation because these are the essential artifacts guiding a software project. Requirements capture stakeholders needs and provide a joint understanding of what features are expected from the software. Architecture documentation lists design principles, patterns and components comprising the software. Documentation of test cases supports testing activities and provides means for quality assurance~\cite{selic2009agile}. 

Only 1\% (7 out of 84) of start-ups in our sample have explicitly stated that 
they do not document requirements in any way. The most popular forms of 
documenting requirements are informal notes and drawings (50 out of 86, 58\%), followed by organized lists (20 out of 86, 20\%).

Responses from the whole sample show that start-ups have some amount of 
documentation debt ($Median = 2.0$), see Fig.~\ref{fig_dimensions}. Exploring 
what start-up characteristics have an effect on the estimates, see 
Table~\ref{table_characteristics_dimensions}, we found that start-ups who are 
active estimate documentation debt lower than acquired or closed companies. We 
also found that teams with sufficient domain knowledge estimate documentation 
debt lower than teams with many gaps in their domain knowledge.



\subsubsection{Architecture debt}

Architecture debt refers to compromises in internal qualities of the software 
such as maintainability, scalability, and evolvability~\cite{Li2015}. 

\begin{figure}[ht]
    \centering
    \includegraphics[width=\textwidth * 4/9]{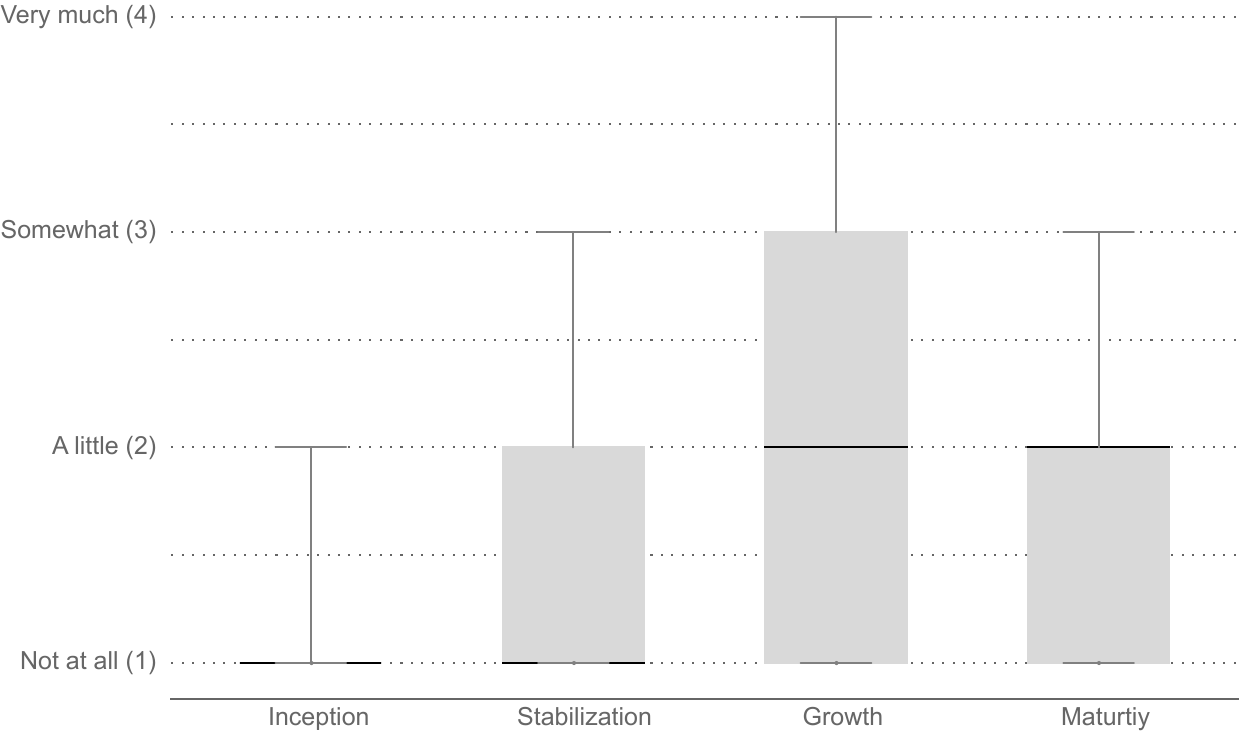}
    \caption{Box-plot showing how in different product phases start-ups estimate architecture debt }
    \label{fig_archi_phases}
\end{figure}

Results from the whole sample show that start-ups experience some 
architectural debt ($Median = 2.0$), see Fig.~\ref{fig_dimensions}. By looking 
into what start-up characteristics have an effect on how respondents estimate 
architecture debt, we found that state of the start-up and the product phase 
have an effect on the estimates, see 
Table~\ref{table_characteristics_dimensions}. Active start-ups have provided 
substantially lower estimates than acquired and closed companies. We found that 
start-ups who have just started on product engineering and haven't yet released 
it to market, experience almost no architectural debt. During stabilization and 
growth phases the estimates become more critical. However, during the maturity 
phase estimates become slightly more optimistic, see 
Fig.~\ref{fig_archi_phases}. Box-plots in the figure show responses from all statements pertaining to architecture debt.

\subsubsection{Code debt}
Code debt refers to a poorly written code. Signs of a poorly written code are, for example, unnecessary complex logic, code clones, and bad coding style affecting code readability. Poorly written code is difficult to understand and change~\cite{Mantyla2003,Tom2013, Palomba2014}.

Results from the whole sample show that start-ups experience some code debt 
($Median = 2.0$), see Fig.~\ref{fig_dimensions}. By looking into what start-up 
characteristics have an effect on how respondents estimate code debt we found 
that state of the start-up, team size, level of per-customer tailoring has an 
effect on the estimates, see Table~\ref{table_characteristics_dimensions}. 
Active start-ups estimate code debt lower than acquired start-ups. Start-ups 
with larger teams (9 or more people), provide higher estimates on code debt 
than small teams. Start-ups who do not offer per-customer customization 
estimate code debt lower. 
However, start-ups that occasionally tailor their product to needs of a 
specific customer estimates their code debt higher.




\subsubsection{Testing debt}
Testing debt refers to lack of test automation leading to the need of manually 
retesting the software before a release. The effort of manual regression 
testing grows exponentially with the number of features, slowing down release cycles and 
making defect detection a time consuming and tedious task~\cite{Tom2013}.

Answers to questions inquiring use of automated testing show that about a third 
(26 out of 86, 30\%) of start-ups are attempting to implement automated testing, and only 17 start-ups (20\%) have explicitly stated that no test automation is used. 

Despite attempts to automate, companies across the whole sample estimate their testing debt somewhat high $(Median=3)$, see Fig.~\ref{fig_dimensions}. Manual exploratory testing is reported as the primary testing practice, regardless of start-up life-cycle phase, team size and engineering experience, and length of operation.
 
\begin{framed}
\textbf{Finding 2:} Despite attempts to automate, manual testing is still the primary practice to ensure that the product is defect free.
\end{framed}

Similar results, showing that only a small number of mobile application projects have any significant code coverage by automated tests, and listing time constraints as the top challenge for adopting automated testing, were obtained by Kochhar et al.~\cite{kochhar2015understanding}.

\subsection{Precedents for technical debt}\label{sec:prectd}

We asked the respondents to estimate various precedents of technical debt in 
their start-ups, such as attitudes towards good software engineering practices, 
pragmatic decisions to make shortcuts in product engineering, communication 
issues in the team, level of team engineering skills, time and resource 
shortages, and lack of established SE processes.

\begin{figure}[ht]
    \centering
    \includegraphics[width=\textwidth * 4/9]{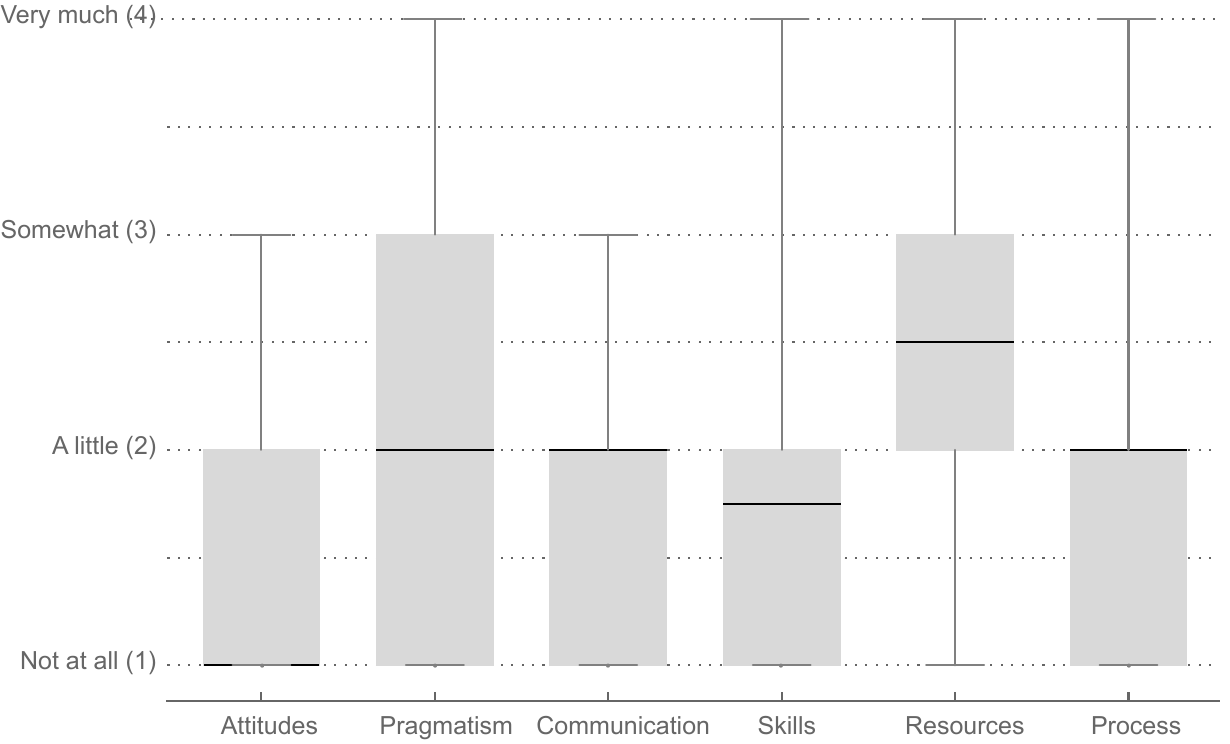}
    \caption{Box-plot showing how the sample estimates different precedents for technical debt}
    \label{fig_precedents}
\end{figure}

Box-plots with median responses from the whole sample are shown in 
Fig.~\ref{fig_precedents}. Higher values indicate stronger agreement with the 
presence of a precedent in the start-up. Poor attitude is the least common 
precedent for technical debt, while resource shortage is 
estimated as the most prevalent precedent.

Looking into what start-up characteristics influence the responses, we find 
that start-up team size and team's engineering skills have a significant effect 
on the estimates overall, see Table~\ref{table_characteristics_precedents}. 
Larger teams of 9 or more people estimate the precedents for technical debt 
higher than small teams. 

In the results we show only characteristics with statistically relevant associations, thus listed characteristics differ between Tables~\ref{table_characteristics_dimensions} and \ref{table_characteristics_precedents}.

\begin{framed}
\textbf{Finding 3:} Start-up team size and level of engineering skills have a significant effect on 
how severe the other precedents are estimated.
\end{framed}

\begin{table*}[!t]
\renewcommand{\arraystretch}{1.1}
\caption{Results of Cramer's V test on associations between precedents and start-up characteristics with $p \less 0.05$ }
\label{table_characteristics_precedents}
\centering

\begin{tabular}{
C{0.2in}L{1.4in}    L{0.5in}    L{0.5in}      L{0.9in}               L{0.6in}            L{0.6in}    L{0.5in}  L{0.3in}}

\# & Characteristic & Attitudes & Pragmatism  & Communication issues &  Skills shortages & Resources shortages & Process & All \\
\hline
1 & State of the start-up             & 0.339  & -      & -      & 0.285 & -     & -      & -      \\
2 & Per-customer tailoring            & -      & -      & -      & -     & 0.345 & -      & -      \\
3 & Overall team size                 & -      & -      & -      & 0.360 & 0.344 & 0.375  & 0.386 \\
4 & Development team size             & -      & 0.432  & 0.387  & 0.429 & -     & 0.403  & -      \\
5 & Level of engineering skills       & -      & -      & 0.331  & 0.465 & -     & -      & 0.377 \\
6 & Level of domain knowledge         & -      & -      & -      & -     & 0.324 & -      & -      \\
7 & Practical experience              & -      & -      & -      & 0.329 & -     & -      & -      \\

\hline
\end{tabular}
\end{table*}

\subsubsection{Attitude towards good engineering practices} 

The responses to questions about following good engineering practices suggest 
that start-up engineers do realize the importance of following good architecture, 
coding and testing practices ($Median=1)$, see Fig.~\ref{fig_precedents}.

Comparing how responses on attitude differ by start-up characteristics we found that start-ups who are active, estimate their attitudes more 
optimistically than acquired or closed companies. 
That is, they agree more with benefits from following good engineering practices, such as coding conventions and throughout testing of the product.

\subsubsection{Pragmatism}
Estimates on statements about pragmatic considerations, that is, prioritization 
of time-to-market over good engineering practices, show that start-ups are 
ready to make shortcuts to speed up time-to-market ($Median=2)$. However, the 
spread of estimates suggests that different companies have very different 
attitudes towards deliberately introducing technical debt, see 
Fig.~\ref{fig_precedents}. Comparing how estimates on attitude differ by 
start-up characteristics we found that start-ups with larger teams of 15 or 
more developers estimate pragmatic precedents higher than smaller teams.

\subsubsection{Communication}
Estimates on statements about communication show that communication issues could be one of the precedents for introducing technical debt, see Fig.~\ref{fig_precedents}. We observe that communication issues become significantly more severe in larger engineering teams of 13 and more people than in smaller teams. Moreover, the results suggest that teams with better engineering skills experience fewer communication issues.

\subsubsection{Engineering skills}
Estimates to what extent start-ups face lack of engineering skills show that skills shortage could be a precedent for accumulating technical debt, see Fig.~\ref{fig_precedents}. 

Comparing how estimates on skill shortages differ by start-up characteristics 
we found that the state of the start-up, team size, length of practical 
experience, 
and level of estimated engineering skills have the significant influence on the 
estimates. 
We found that active start-ups estimate skills shortages 
lower than closed down companies. A somewhat expected result is that teams with 
adequate engineering skills provide significantly lower estimates for 
challenges associated with skills shortages.

\subsubsection{Resources}
Looking at differences between estimates on time and other types of resources we found that they are tied together, see 
Fig.~\ref{fig_precedents}. That is, companies reporting time shortages also report resource shortages. A potential explanation for the association is that time pressure is created internally by a need to get the product out and start generating revenue, and not by an external market pressure. We also found that per-customer customization, overall team size, and level of domain knowledge has an effect on how start-ups estimate resource shortages.

Estimates of resources and time shortages show that resource issues are the 
highest estimated precedent for technical debt ($Median = 2.5$).
We find that occasional per-customer tailoring is associated with higher 
estimates on resource shortages. Potentially, start-ups suffering from lack of 
resources opt for occasional customization to serve needs of an important 
customer, thus acquiring resources for further development. Start-ups with 
smaller teams of 1-3 people estimate resource shortages lower than larger 
start-ups of 9-12 people. A plausible explanation for this association could be 
that supporting a larger team requires more resources.

\subsubsection{Process}
Respondent estimates on the software engineering process issues show that 
frequent and unplanned changes occur and could cause difficulties in avoiding 
technical debt, see Fig.~\ref{fig_precedents}. We found that estimates on 
process issues become more severe as team size grows.

\subsection{Outcomes of technical debt}\label{sec:outtd}

To explore potential outcomes of technical debt we presented respondents with statements exploring to what extent team productivity and product quality are concerns in their start-ups. Estimates from the whole sample show that start-ups experience some quality and 
productivity issues ($Median = 2$) that could be associated with accumulated technical debt. We found that team size is the only characteristic that influences the estimates ($Cramer's V = 0.362$). 

Looking into what types of technical debt are associated with specific outcomes linked to technical debt, we found a clear association between estimates of technical debt and estimates of the outcomes, see Table~\ref{table_evidence_outcome}.

Code debt has the most severe impact on both productivity and quality. Architecture debt have a similar effect, albeit to a lesser extent. Documentation debt impairs productivity. However, we did not find a statistically significant association between testing debt and loss of productivity or quality.

\begin{framed}
\textbf{Finding 4:} We found that from all types of technical debt, code debt have the strongest association with productivity and quality issues. 
\end{framed}


\begin{table}[!t]
\renewcommand{\arraystretch}{1.1}
\caption{Results of Cramer's V test on associations between types of technical debt and outcomes with $p \less 0.05$ }
\label{table_evidence_outcome}
\centering

\begin{tabular}{
C{0.1in}L{0.8in}    L{0.5in}    L{0.7in}      L{0.5in}}

\#      & Debt type & \multicolumn{3}{c}{Impact on:}\\
&          & Quality & Productivity  & Both  \\
\hline
1 & Documentation   & -       & 0.332   & 0.344 \\
2 & Architecture    & 0.399   & 0.331   & 0.440 \\
3 & Code            & 0.445   & 0.471   & 0.532 \\
4 & Testing         & -       & -       & -     \\
\hline
  & All types       & 0.363   & 0.459   & 0.463 \\

\end{tabular}
\end{table}






\section{Discussion} 
\label{sec_dis}

\subsection{Reflections on the research questions}


Our results on how start-ups estimate technical debt show that active start-ups 
estimate aspects of technical debt significantly lower than closed or acquired 
start-ups, see Finding~1 in Section~\ref{sec:dimtd}. 
A plausible explanation for this result 
could be that lower technical debt helps start-ups to have a more stable and easier to maintain product. 
Thus giving a start-up more room for evolving the product into something the 
market wants, i.e. to pivot~\cite{bajwa2016software}. However, excess technical 
debt hinders product evolution and could be one of contributing factors to the 
shutdown of a company.

An alternative explanation is that technical debt could be invisible and compensated
by the team's implicit knowledge. However, when a start-up is acquired by 
another company and the product is transferred to another team, all the technical 
debt becomes visible. Difficulties to capture undocumented knowledge and the 
associated drop in performance of the receiving team has been recognized in the 
context of agile project handover~\cite{stettina2013there}.

Results on how different types of technical debt are estimated show that the 
most technical debt is estimated in the testing category, even though start-ups 
do attempt to automate tests, see Finding~2 in Section~\ref{sec:dimtd}. A 
potential explanation could be that start-ups lack certain prerequisites for 
full implementation of automated testing~\cite{Voas}. Excess technical debt in 
other categories, for instance, difficult to test code, lack of requirements 
documentation, and an unclear return of investment, could be hindering the 
implementation of test automation, as it is also observed in traditional, more 
mature companies~\cite{rice2003surviving, joorabchi2013real,kochhar2015understanding}. However, we could not find any statistically significant association between testing debt and quality or productivity issues.

The comparison of results on architecture debt from start-ups in different 
life-cycle phases shows another interesting pattern.
Start-ups who have not yet released their products to market experience very little architecture debt, the debt increases as the product is delivered to the first customer and peaks at the growth stage when start-ups focus on marketing the product and drops slightly as start-ups mature, see Fig~\ref{fig_archi_phases}. Marketing of the product could be a source of new challenges for the product development team. For example, the product must support different configurations for different customer segments, provide a level of service, and cope with a flow of requests for unanticipated features~\cite{Crowne2002, Dahlstedt}. Earlier shortcuts in product architecture are therefore exposed and must be addressed.

Overall team size and level of engineering skills could be the most important 
characteristics contributing to precedents and linked to technical debt in most 
dimensions, see Finding~3 in Section~\ref{sec:prectd}. Larger teams of 9 or 
more people experience more challenges and report higher technical debt in all 
categories. This finding is similar to Melo et al.~\cite{DeMelo2013} studying 
productivity in agile teams. Smaller teams are better aligned and more 
efficient in collaboration with little overhead. However, as the team size 
grows more processes and artifacts for coordination are 
needed~\cite{staats2012team}. Therefore larger teams have more artifacts that 
can degrade.

Team size could be an indicator of the general complexity of a start-up and the product. More people are added to the team when there are more things to be taken care of. Therefore, the technical debt could stem not only from the number of people but also from increasing complexity of the organization itself.


Our results show that increase in team size is also associated with outcomes of 
technical debt, a decrease in productivity and product quality, see Finding~4 
in Section~\ref{sec:outtd}.
This result could be explained by our earlier discussion on how larger teams require more coordination for collaboration. However, the more critical estimates by larger teams could be also associated with the increase in product complexity as new features are added. Rushing to release new features could contribute to the accumulation of technical debt until deliberate, corrective actions are taken, as observed in mobile application development~\cite{hecht2015tracking}. 

As a software product grows, it naturally becomes more difficult to maintain. 
For instance, if individual product components do not change and the new 
components are at the same quality level as existing ones, the increased number 
of components and their dependencies requires more effort from engineers to 
maintain the product and creates more room for defects~\cite{Izurieta2013}. 
This is software decay and is not the same as avoidable technical debt stemming 
from the trade-off between quality and speed. Distinguishing between true 
technical debt and software decay is an important next step in providing 
practical support for software-intensive product engineering in start-ups.

\subsection{Implications for practitioners}

This study presents several implications for practitioners:

\begin{enumerate}

\item Start-up teams with higher level of engineering skills and respondents with more experience perceive aspects of technical debt more severely. Less skilled teams may not be aware of their practices introducing additional technical debt, and amount of technical debt in their products. Using tools and occasional external expert help could help to identify unrealized technical debt, and to improve any sub-optimal practices.

\item Start-up team size correlates with more severe precedents and outcomes of 
technical debt. Keeping a team small and skilled could be a strategy to 
mitigate precedents for technical debt. To support growth of the team, more coordination practices need to be introduced, and impact on technical debt monitored. Additional coordination practices require more maintenance of coordination artifacts. Thus, there is a practical limit how large a team can grow before it needs to be divided into sub-teams.

\item There is an association between levels of technical debt and a start-up 
outcome. Having less technical debt could give a start-up more room for 
pivoting and product evolution in the long term. 

\item There are certain moments when the effects of technical debt are the most 
severe. For example, shipping a product to a large number of customers, scaling 
up the team, and handing the product over to another team. The anticipation of 
such moments and adequate preparations could help to mitigate the negative 
effects of technical debt.

\item The most significant type of technical debt in start-ups is code smells. We found that poorly structured and documented code has the strongest association with issues in team productivity and product quality. However, detection of code smells can be automated with open-source tools, thus alleviating removal of this type of debt.

\end{enumerate}

\section{Conclusions and future work}
\label{sec_conclusions}
In this paper, we report how technical debt is estimated in start-ups building 
software-intensive products. We explore to what extent precedents, dimensions, and outcomes, identified by earlier studies, are relevant in the start-up 
context. We attempt to identify what start-up characteristics have an 
amplifying or remedying effect on technical debt.

Our results show that, even though start-up engineers realize the importance of 
good 
engineering practices, they cut corners in product engineering, mostly due to 
resource pressure and a need for faster time to market. 
The results suggest that precedents for technical debt become more severe as 
start-ups evolve and severity of the precedents 
could be associated with the number of people working in a start-up and a product 
life-cycle phase. 

Our results show significantly different estimates from closed, acquired and operational start-ups. The differences highlight how start-ups use technical debt as a leverage, and emphasizes the importance of careful technical debt management.


This exploratory study leads to a formulation of several hypotheses:

\begin{enumerate}[label=(\alph*)]
    \item Technical debt peaks at the growth stage when a start-up attempts to 
    market the product. 
    \item The number of people in a team amplifies precedents for technical 
    debt.
    \item There is an association between a start-up outcome and their 
    technical debt management strategy.
\end{enumerate}

We aim to explore these hypotheses further by triangulating results from this study with qualitative data from
interviews and artifact analysis.


\section{Acknowledgments}
The authors of this paper would like to thank all practitioners who found time 
and motivation share their experiences. Reaching this diverse population of 
start-ups would not be possible without help and support from Software Start-up 
Research Network community.
and specifically Nana Assyne, Anh Nguyen Duc, Ronald Jabangwe, Jorge Melegati, Bajwa Sohaib Shahid, Xiaofeng Wang, Rafael Matone Chanin, and Pekka Abrahamsson.

Work of R. Prikladnicki is supported by Fapergs (process 17/2551-0001205-4).

\bibliographystyle{ACM-Reference-Format}

\bibliography{library.bib}{}

\end{document}